# Material Optimization for Fermi Surface Shape Control of Tl-based Cuprate Superconductors

Satoaki Miyao[a], Hirofumi Sakakibara[b], Isao Maruyama[c], Kazuhiko Kuroki[d], and Koichi Kusakabe[a,*]

[a]*Department of Materials Engineering Science, Osaka University, 1-3 Machikaneyama-cho, Toyonaka, Osaka 560-8531, Japan*

[b]*Department of Engineering Science, University of Electro-Communications, 1-5-1 Chofugaoka, Chofu, Tokyo 182-8585, Japan*

[c]*Department of Information and Systems Engineering, Fukuoka Institute of Technology, 3-30-1 Wajiro-higashi, Higashi-ku, Fukuoka 811-0295, Japan*

[d]*Department of Physics, Osaka University, 1-1 Machikaneyama-cho, Toyonaka, Osaka 560-0043, Japan*

*Corresponding author. Tel/Fax: +81 6 6850 6406.
E-mail: kabe@mp.es.osaka-u.ac.jp (K. Kusakabe)

## ABSTRACT

**To show an optimization method of element substitution for cuprate superconductors, we investigate Fermi surface shape of $TlR_2A_2Cu_3O_9$ with R=La, Y, and A=Li, Na, K, Rb, Cs. We adopt the generalized gradient approximation in the density-functional theory (DFT-GGA) for the study of over-doped phases of these unknown cuprates. The electronic structures of crystals optimized by DFT-GGA show systematic element dependence in a Fermi surface shape controlling parameter, *r*, of Cu $dx^2$-$y^2$ bands, where nearly absent $dz^2$ component at the Fermi level and smaller *r* keeping $t_1$ suggest enhancement of the superconducting transition temperature within the spin-fluctuation mechanism. For $TlYRb_2Cu_3O_9$, smaller *r* by a reduction factor larger than 10% compared to a reference system of $TlBa_2Ca_2Cu_3O_9$ (TBCCO) appears in the outer $CuO_2$ plane, but with 12 % reduction in $t_1$. For $TlR_2Li_2Cu_3O_9$, (R=Y, La), smaller *r* by a factor larger than 1% appears keeping $t_1$ as large as TBCCO in the inner plane. Our method may be used to predict an optimized material structure referencing known cuprate superconductors.**

# 1. INTRODUCTION

In a series of studies of the Cu $dx^2$-$y^2$ bands of cuprates,[1-6] two of the present authors and their coworkers have deduced relevant factors decisive for the high-$T_c$ paying attention to the nesting property of the Fermi surface. At first, the Fermi surface shape only is not enough to guarantee the high $T_c$. It is requested to erase character of Cu $dz^2$ orbital from the Fermi level.[1-3] Keeping this condition, a good nesting is to be sought in the Fermi surface of the Cu $dx^2$-$y^2$ band.[4-6] When we adopt a tight-binding description with $t_1$, $t_2$, and $t_3$, which denote nearest, next-nearest, and third nearest hopping amplitudes among the Cu $dx^2$-$y^2$ Wannier orbitals, respectively, the shape of the Fermi surface is effectively described by $r=|t_2/t_1|+|t_3/t_1|$.[2-5] We call $r$ "the Fermi surface shape controlling parameter", which is abbreviated as the shape controlling parameter from now on.

When $r$ becomes smaller, the better Fermi surface nesting is achieved. It is further explored that relative positions between $dx^2$-$y^2$ and 4s orbitals of Cu on the energy axis works to change $r$.[4-6] This is because orbitals with Cu 4s character may mediate long range transfers $t_2$ and $t_3$ for $dx^2$-$y^2$ orbitals. The lesser 4s contribution becomes, the more preferable Fermi surface shape of a single-band model on a square lattice appears. When $r$ is getting small, keeping $t_1$ as large as or larger than that of a reference system, an ideal electronic structure in an effective pure 3d band is obtained starting from the reference. Owing to these rules, the authors have called the $T_c$ enhancement "the orbital distillation effect".[4-6]

The value of $r$ may be reduced by the external pressure.[4,5] In addition, we might be able to find material solutions, namely unknown phase of materials, in order to decrease $r$ to obtain higher $T_c$ than ever found. Although the guiding principle suggests less 4s character in the $dx^2$-$y^2$ band to improve the orbital distillation effect, 4s orbitals of Cu inevitably exist in real materials. Thus, it is by no means trivial whether we can control the Fermi surface shape by element substitution. Therefore, it is important to investigate material dependence of $r$.

Our purpose is to show that $r$ is changeable and controllable by a quantitatively predominant amount. Optimizing chemical composition in a typical multi-layered Tl-1223 compound, *i.e.* $TlBa_2Ca_2Cu_3O_9$ (TBCCO) [7-9] by substituting elements, we will show that the shape controlling parameter can be reduced by as much as 10 % from the reference. TBCCO has three $CuO_2$ planes. By looking at the inner layer and the outer layer independently, two conditions on $t_1$ and $r$ can be refined at the same time in some examples.

The organization of the paper is the followings. In Sec. II, we introduce our strategy. We will explain a reason why we selected TBCCO as a reference system. In Sec. III, numerical data on the tight-binding parameters and the shape controlling parameter are determined for some unknown structures. The structures are given by substituting some elements in TBCCO by the others. Crystal structure and the electronic band structures are optimized theoretically. In Sec IV, we will analyze origin of the parameter change. We conclude our research on the material optimization method in Sec V.

# 2. METHODS

## 2.1 DFT band structure calculation

To show a tunable range of *r*, we consider both known and unknown material structures. We need to know electronic structures of some unknown crystals. Thus, we are required to determine crystal structures theoretically.

We adopt simulation techniques realized by the density functional theory.[10,11] We will show that the crystal structure of TBCCO is reproduced with reasonable accuracy by the generalized gradient approximation (GGA). It is known that the electronic structures with well-developed large Fermi surfaces in cuprate are reproduced even by a DFT energy functional.[12-16] As for the over-doped cuprates, therefore, we safely start our discussion starting from the GGA calculations.

Calculation details are as follows. Density functional theory modeling of materials is performed. We adopt a GGA energy functional using the Perdew-Burke-Ernzerhof form of the exchange correlation energy[17] as implemented in the VASP code.[18,19] The projector-augmented-wave method (PAW) is used.[20] The cut-off energy for the expansion of wave functions was set to 500 eV (36.75 Ry) to calculate the optimized geometries and to determine the band structure. The Brillouin zone integration to obtain the self-consistent charge density was performed using a uniform $8\times8\times4$ k-point grid including the $\Gamma$ point. All atoms were fully relaxed until the forces were less than $1.5\times10^{-2}$ eV/Å ($6\times10^{-4}$Ry/a.u.).

Reliability of structure determination of cuprates was tested by applying it to TBCCO. Determined crystal parameters are compared with observed values in Table I. We performed both inner coordinate optimization with a fixed unit cell, and full optimization of the crystal structure. In the former, lattice constants found in an experiment[7-9] were used for the tetragonal unit cell. For the latter simulation, the constant pressure optimization scheme was used. After the full optimization, the pressure was less than 0.85[kbar]. The a axis was almost unchanged within 0.26% elongation in the full optimization, while the c axis became 1.9% longer than the observed value. In both of the optimization, in-plane Cu-O-Cu angle in the outer plane becomes about $170^{\circ}$, which is rather shifted from the observation result of $176^{\circ}$. Therefore, we may say that DFT-GGA shows us stronger bending in each Cu-O-Cu bond of the outer plane. Owing to the symmetry, the inner plane keeps its flatness. The Cu-O bond length for the apical O at the outer plane became 2.68Å. The Cu-Cu distance giving the inter-plane distance was 3.30, which is 3% longer than the natural crystal value.

Our simulation data reproduces characteristic features of the former DFT band calculations.[15] In Fig. 1, we show the Kohn-Sham electronic band structure and the Fermi surface for the optimized crystal structure. On the dispersion curve in Fig. 1 (b), we indicate the projected density of states (p-DOS) by colored symbols. The blue diamonds represent weights on Cu $dx^2$-$y^2$, while the green squares are for Cu $dz^2$. We can clearly see that $dz^2$ contribution is almost invisible around the Fermi energy, which is at the origin of the vertical axis. TBCCO has three Fermi surfaces in rod-like shape surrounding the M point. The multi Fermi surfaces originate from three $CuO_2$ layers in a unit cell. The surfaces are not in square poles but rather close to columnar. TBCCO obtained here is used as a reference material to analyze designed materials in the following sections.

### 2.2 Wannier transformation

To evaluate the transfer parameters of cuprates, we perform the Wannier transformations. The Wannier90 code[21] is used to optimize the maximally localized Wannier functions.[22] In the following sections, we analyze transfer parameters obtained as the Fourier transformation of the DFT-GGA spectrum selected by the disentanglement procedure.[23] Since the transfer parameters depend on a modeling, we select symmetry for projection operators to derive a reasonable model.

Here, TBCCO is a triple layered cuprate. We apply a six-band model, which consists of Cu $dx^2$-$y^2$ and $dz^2$ orbitals, to fit the electronic band structure around the Fermi level. By this selection, we are able to reconfirm irrelevance of Cu $dz^2$ orbitals for the Fermi surface. Test calculations of TBCCO actually show the good fitting of the DFT-GGA band structures. Note that the Wannier orbitals consist of Cu 3d orbitals hybridized with O 2p orbitals. The DFT-GGA branch with high contributions of Cu $dx^2$-$y^2$ and $dz^2$ are actually extracted out. The same procedure to derive a six-band model is adopted for materials structures considered in Sec III.

### 2.3 Material design

As shown in Sec. II B, our reference material, TBCCO, shows imperfect Fermi surface shape, where nesting property is not so strong. The orbital distillation effect is expected to be refined by changing materials parameters. Guided by knowledge on the effect of Cu 4s orbitals,[4-6] we now consider effective substitution of elements in TBCCO. Here we propose our strategies for the materials design modifying the multi-layered cuprate. A key insight is that the effective potential for Cu 4s orbitals is affected by surrounding ionic potentials.

Substitution of atoms would cause a change in the effective potential and the band structure. We want to introduce a change in the energy levels strong enough to affect Cu 4s levels, keeping good properties of $CuO_2$ planes as untouched as possible. By some test calculations for TBCCO, we found that substitution of Ba or Ca by other di-valent alkaline-earth elements is not so remarkable. So, we consider wider classes of elements, which may cause possible mono-valent or tri-valent cations in the crystal. For the replacement, it is better to have several atomic sites in a unit cell. Indeed, to keep a good flat $CuO_2$ plane for a single-layered cuprate, after the replacement of Ba by the others, we would be requested to consider a super cell. Thus, a single-layered material like $HgBa_2CuO_4$ is not appropriate for simulations compared to multi-layered cuprates, especially three-layered material whose inner $CuO_2$ plane is kept flat by the symmetry.

In some Tl-based cuprates, over-doped phase with good metallic band structures can appear with good integer-valued stoichiometric ratios of elements. Actually, a simulation of metallic $TlBa_2Ca_2Cu_3O_9$ is possible using a primitive unit cell, whereas we need to introduce approximate model crystals for a conventional set up of Hg-based cuprates. Therefore, the selection of Tl-compounds has an advantage.

Owing to the above reasons, we choose TBCCO as a starting reference. Replacement of divalent cations by mono- and tri-valent cations is considered. We do not intend to introduce extra magnetism. Thus, we choose La and Y for tri-valent elements and alkali metal elements (Li, Na, K, Rb, and Cs) for mono-valent cations. To keep the charge balance, Ba is replaced by La or Y, while Ca is replaced by one of alkali-metal elements. When we compute a compound with La at Ca sites and K at Ba sites, the total energy is higher than that with La replacing Ba, and K replacing Ca. This is because the Ba site has higher coordination of oxygen atoms than the Ca site. Therefore, tri-valent elements are kept in Ba sites, while Ca sites are substituted by alkali atoms.

## 3. RESULTS

### 3.1 Structure Optimization

We performed structural optimization simulations and obtained estimated crystal structures for 10 combinations of ingredients, $TlR_2A_2Cu_3O_9$ with R=La, Y, and A=Li, Na, K, Rb, Cs. Owing to less accuracy in evaluation of the internal stress tensor $\sigma_{\alpha\beta}$ than the inter-atomic forces, we fixed the size of the tetrahedral unit cell. The cell parameters were those for TBCCO. The result obtained by this fixed-cell approximation may be regarded as a simulation in a uniaxial stress. We did confirm that evaluated structures showed finite internal stress, whose diagonal elements were in a range of [-90, 180] (kbar), except for La-Cs and Y-Cs compounds, in which $\sigma_{\alpha\alpha}$ was (116,116,269) (kbar) and (52,52,224) (kbar), respectively with zero off-diagonal elements. For a crystal containing elements with small ionic radii, the cell tends to shrink, while a structure having large ions tends to expand. We expect errors in the stress tensor in a few tens of kbar.

As an example, we show optimized crystal structure of $TlLa_2Li_2Cu_3O_9$ in Figure 2. There are several structural parameters which directly affect the transfer parameters of electrons in the $CuO_2$ plane. In this section, we consider 1) in-plane Cu-O-Cu angle, 2) Cu-Cu distance giving the inter-plane distance and 3) Cu-O bond length for the epical O at the outer plane.

In Table II, we show the structural data. For the materials considered, outer plane is largely distorted to have the Cu-O-Cu angle smaller than $170^{\circ}$. We should note that the angle is redefined from that for TBCCO (Compare Fig. 2 with Fig. 1). This means that in-plane oxygen atoms in the outer plane are attracted to the rare-earth elements. This behavior is owing to large positive charges of the rare-earth ions. The inter plane distance is larger than 3.30 Å, which was found in a calculation of TBCCO. Furthermore, the distance between the apical O and Cu in outer plane is rather short. These structural features and its relevance for the transfer parameters would be discussed in elsewhere.

### 3.2 Electronic Band Structure

All compounds considered possess three Fermi surfaces of Cu $dx^2$-$y^2$orbitals. A few additional Fermi surfaces coming from oxygen bands appear in several structures except for compounds with Li. (Fig. 3) These small pockets of oxygen bands locate around the M point. Fermi surface shape for $dx^2$-$y^2$ bands looks similar among 10 candidate materials. Therefore, we will do quantitative evaluation of the Fermi surfaces in the next section.

The p-DOS profile of these materials ensures that Cu $dz_2$ contribution is almost invisible around the Fermi level. This is confirmed in the DFT-GGA band structures. As examples, we show the bandstructure of $TlY_2Rb_2Cu_3O_9$ in Fig. 4. Therefore, we will concentrate on the Fermi surface shape controlling parameter in the next section.

### 3.3 Materials Dependence of the Shape Controlling Parameter

Maximally localized Wannier functions were determined for the low energy $dx^2$-$y^2$ and $dz^2$ bands. The effective tight-binding Hamiltonian was determined for designed materials. Owing to bending of Cu-O-Cu bonds, the absolute value of the nearest neighbor hopping, $t_1$, of $dx^2$-$y^2$ electrons decreased in the outer plane from that of the reference system (TBCCO). (Table III)

To consider Fermi surface shape, normalized intra-plane tight-binding parameters for electron hopping to second-nearest and third-nearest sites, $t_2/t_1$, and $t_3/t_1$, are important factors.[24,25] Obtained $t_2/t_1$ and $t_3/t_1$ for an outer $CuO_2$ plane are varying around those for TBCCO. For the inner plane, however, structures with alkali- and rare-earth atoms have smaller $|t_2/t_1|$ than that for the reference TBCCO, keeping larger $|t_1|$ than the reference.

We show obtained $r$ of the outer and inner $CuO_2$ planes in Fig. 5 and Fig. 6, respectively. In the outer plane, the larger the ionic radius becomes, the smaller $r$ becomes, except for Cs compounds. Among the $TlR_2A_2Cu_3O_9$ series, reduction in r compared to the highest is expected to be more than 50%. When we compare $r$ for designed materials and TBCCO, we have 10% reduction in $TlY_2Rb_2Cu_3O_9$. Note that $t_1$ is -0.4478 (eV) and the effective band width should be 12% smaller than TBCCO in the outer plane.

The result for the inner plane gives us another insight. The value of $r$ does not so drastically change by chemical substitution. However, for $TlR_2Li_2Cu_3O_9$, inner plane transfer shows even smaller $r = |t_2/t_1|+|t_3/t_1|$ than the reference. At the same time, $t_1$ becomes even larger than TBCCO.

## 4. DISCUSSIONS

In the discussion given below, let us assume spin fluctuation mediated pairing in the cuprates. Within this scenario, in the absence of $dz^2$ mixture in the Fermi surface, there can be two parameters that determine $T_c$ (for a fixed band filling)[4-6]. One is $U/t_1$, where the nearest neighbor hopping $t_1$ is directly proportional to the band width, and $U$ is the on-site repulsion, whose material dependence is assumed to be weak. $T_c$ is optimized for a certain optimal value $[U/t_1]_{opt}$, and considering the fact that $T_c$ increases with pressure (i.e., for larger $t_1$), it is likely that most of the cuprates lie in the strongly interacting regime ($U/t_1>[U/t_1]_{opt}$). The second parameter is the shape controlling parameter $r$ as mentioned previously, i.e., smaller $r$ gives higher $T_c$. Thus, if we find a material with smaller $r$ with a similar $t_1$ compared to the reference material, then the material should exhibit enhanced Tc.

Note that this "$r$ vs. $T_c$" rule is the opposite to those discussed in previous studies[24,25]. This is because we are now assuming cases where the $dz^2$ contribution to the Fermi surface is very small; when the $dz^2$ component mixture becomes strong, the warping of the Fermi surface is reduced ($r$ is apparently small), but $T_c$ is suppressed due to the $dz^2$ orbital mixing itself[1-6].

For $TlY_2Rb_2Cu_3O_9$, smaller $r$ by a reduction factor larger than 10% compared to a reference system of $TlBa_2Ca_2Cu_3O_9$ appears in the outer $CuO_2$ plane. Even if we have 12 % reduction in $t_1$, the prominent reduction in $r$ suggests that the system may be a candidate for a high $T_c$ material. For $TlR_2Li_2Cu_3O_9$, (R=Y, La), smaller $r$ by a factor larger than 1% appears keeping $t_1$ as large as TBCCO appears in the inner plane.

Assuming that two-body interactions in these materials are similar to the reference, we conclude that the superconducting transition temperature may be comparable to or even higher than TBCCO in some of the proposed material structures. Since the inter $CuO_2$ plane distance is estimated to be rather large, the inter plane interaction should be carefully evaluated.[26,27] This problem is left for the future study.

## 5. CONCLUSIONS

We have shown that the Fermi surface shape controlling parameter, $r$, is controllable by the chemical substitution. The amount of change in $r$ can create measurable change in $T_c$, when we can create a series of materials.

## Acknowledgements

This work was supported by Grant-in-Aid for Scientific Research (No. 235404080) from MEXT, Japan, and by G-COE, "Core research and engineering of advanced materials-interdisciplinary education center for materials science". The authors thank the Supercomputer Center, Institute for Solid State Physics, University of Tokyo, and the computing system for research, research institute for information technology, Kyushu University, for the use of the facilities.

**Figure captions**

**FIG. 1.** (a) The optimized structure of $TlBa_2Ca_2Cu_3O_9$. (b,c) The Kohn-Sham band structure and (d) the Fermi surface of $TlBa_2Ca_2Cu_3O_9$. θ, *d*, and *l* denote the Cu-O-Cu angle, Cu-Cu distance, and a distance from Cu to an apical oxygen, respectively. The Fermi level is at the origin of the Kohn-Sham orbital energy. In the band structure, the projected density of states is marked by blue diamonds ($dx^2$-$y^2$) in (b) and green squares ($dz^2$) in (c).

**FIG. 2.** .The optimized structure of $TlLa_2Li_2Cu_3O_9$ in a tetragonal unit cell. θ, *d*, and *l* denote the Cu-O-Cu angle, Cu-Cu distance, and a distance from Cu to an apical oxygen, respectively.

**FIG. 3.** The Fermi surface determined for optimized structures of (a) $TlLa_2Li_2Cu_3O_9$, (b) $TlLa_2Na_2Cu_3O_9$, (c) $TlLa_2K_2Cu_3O_9$, (d) $TlLa_2Rb_2Cu_3O_9$, (e) $TlLa_2Cs_2Cu_3O_9$, (f) $TlY_2Li_2Cu_3O_9$, (g) $TlY_2Na_2Cu_3O_9$, (h) $TlY_2K_2Cu_3O_9$, (i) $TlY_2Rb_2Cu_3O_9$, and (j) $TlY_2Cs_2Cu_3O_9$.

**FIG. 4.** The Kohn-Sham band structure of $TlY_2Rb_2Cu_3O_9$. The projected density of states is marked by blue diamonds ($dx^2$-$y^2$) in (a) and green squares ($dz^2$) in (b).

**FIG. 5.** The Fermi surface shape controlling parameter, *r*, determined by transfer parameters at the outer plane of materials in a Tl-1223 form.

**FIG. 6.** The Fermi surface shape controlling parameter, *r*, determined by transfer parameters at the inner plane of materials in a Tl-1223 form.

**Table captions**

**Table 1** Crystal parameters and structure parameters of $TlBa_2Ca_2Cu_3O_9$. Obs. is the experimentally observed values.[7-9] Values by calculations are for a fixed cell simulation (cell fixed) and a cell relaxed simulation (cell opt.).

**Table 2** Structure parameters determined by the structure optimization simulations for materials in a Tl-1223 form. See also the text.

**Table 3** The tight-binding parameters for materials in a Tl-1223 form. See also the text.

**Figures**

**Figure 1** (1.4-column figure for print)

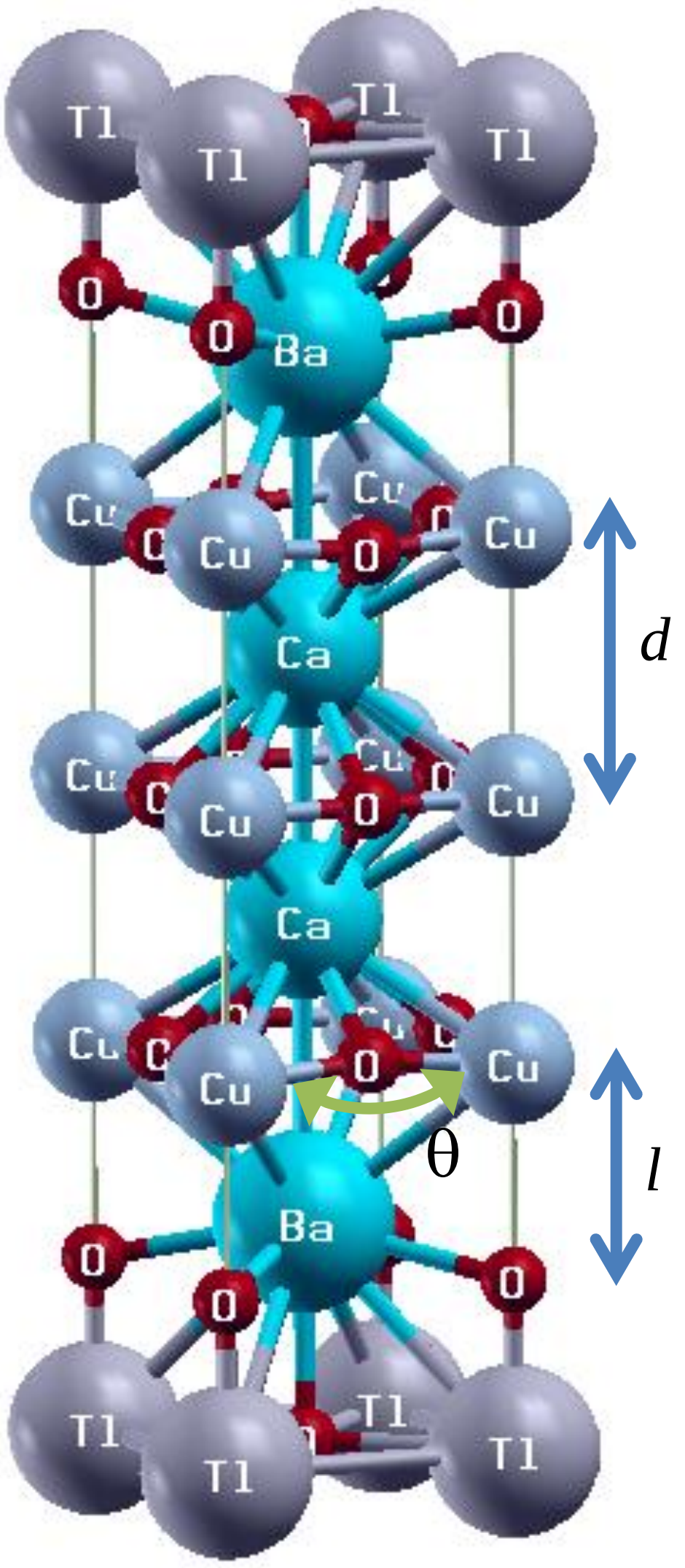

**Figure 1 (a)**

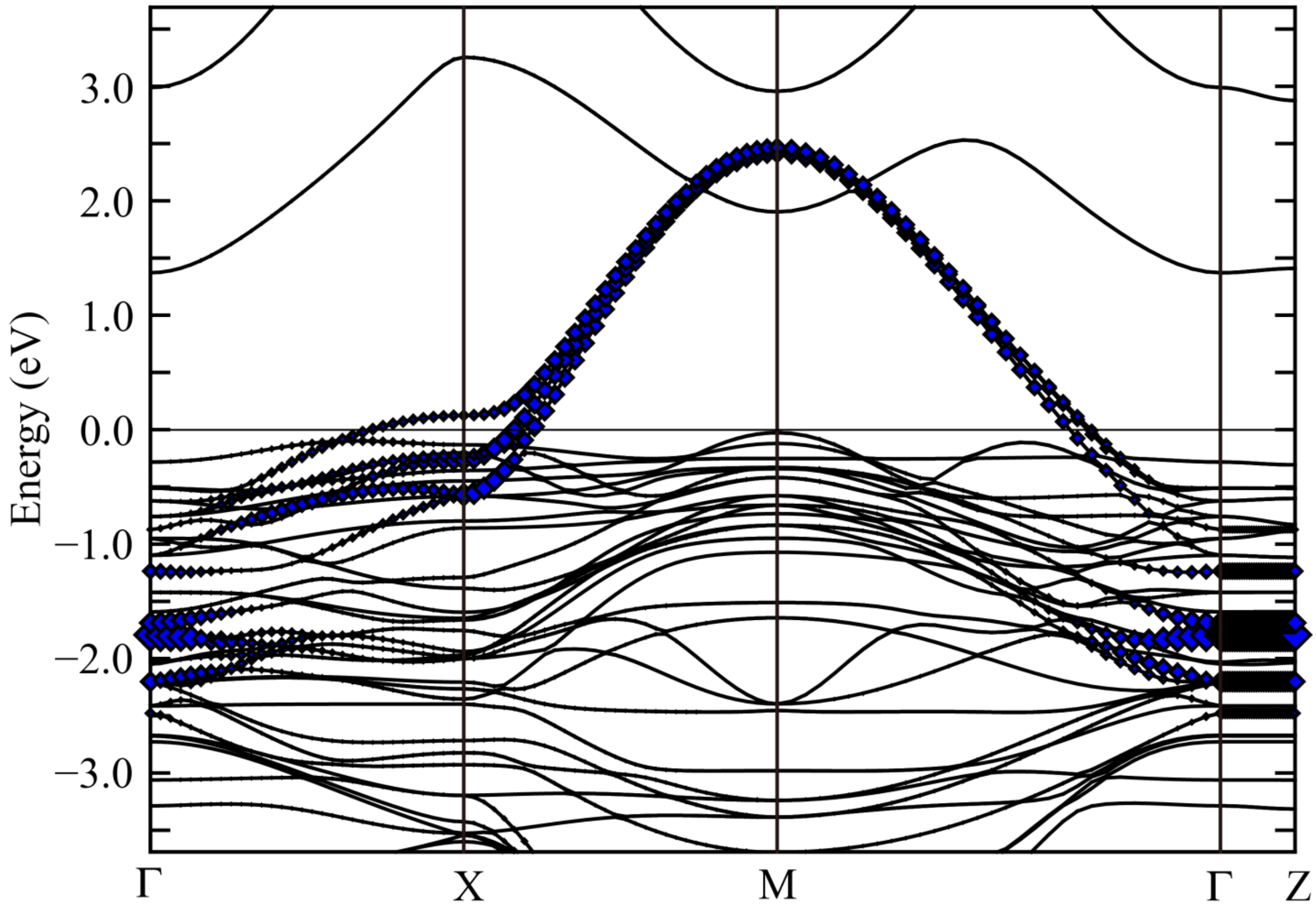

**Figure 1 (b)**

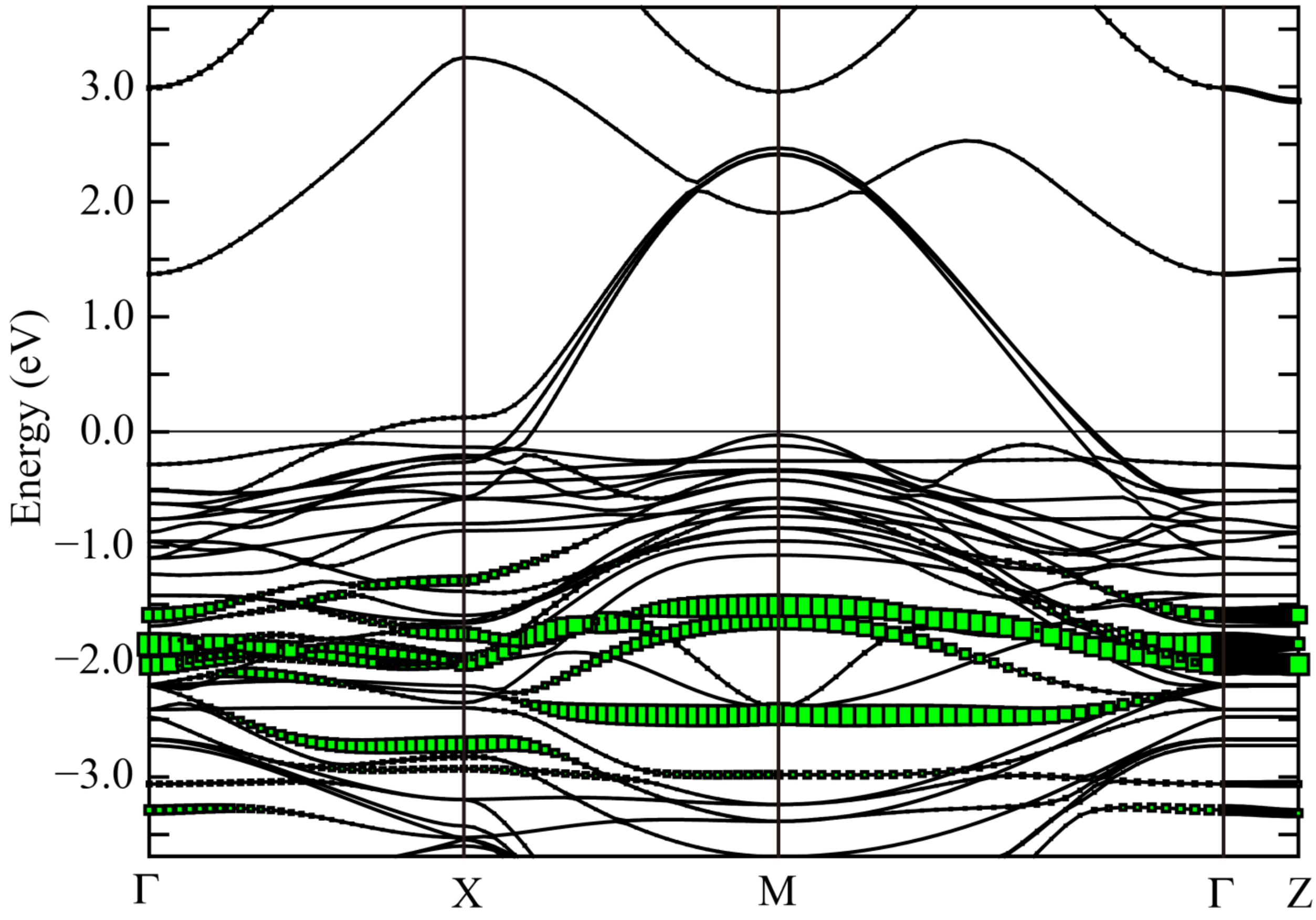

**Figure 1 (c)**

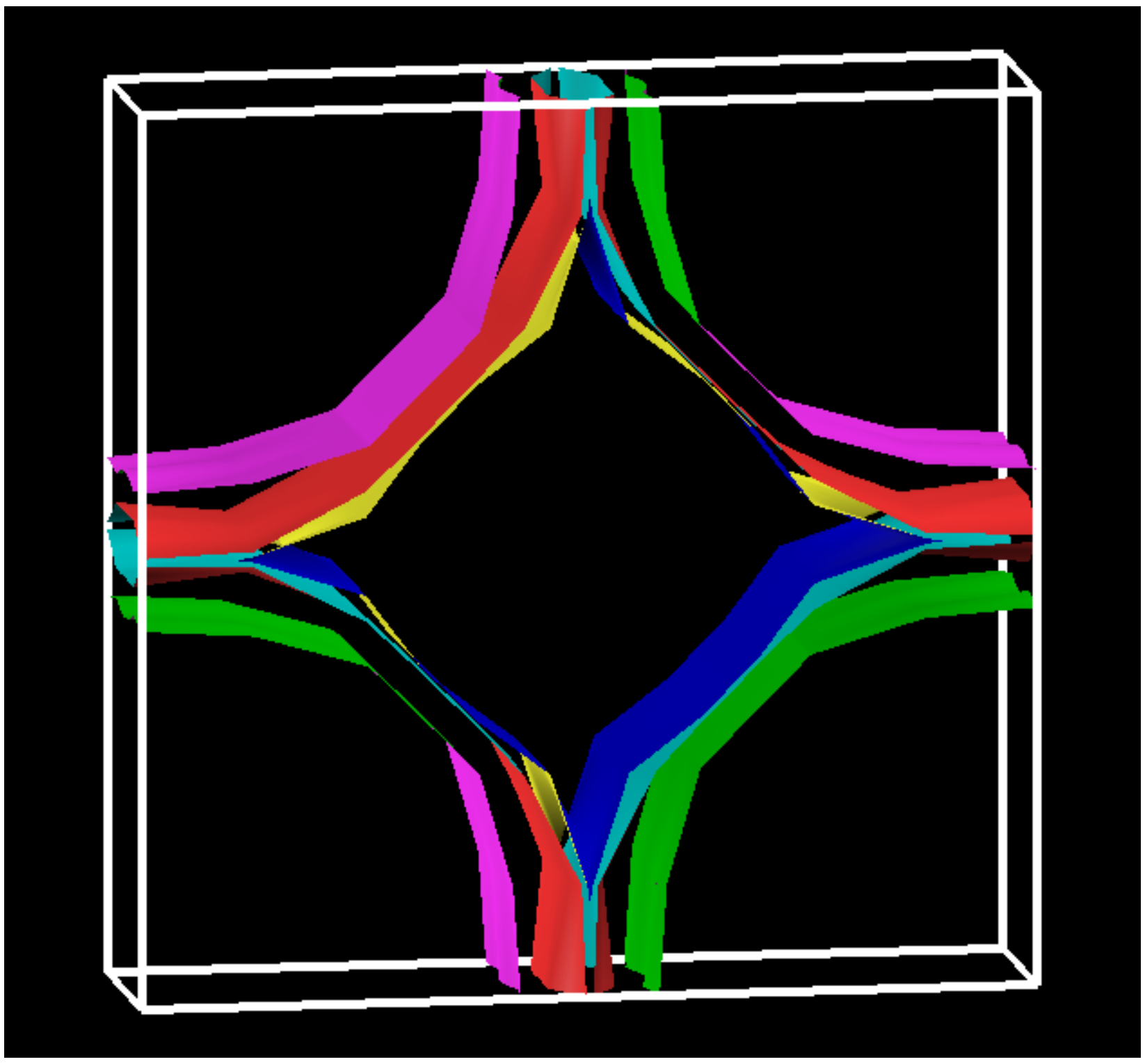

**Figure 1 (d)**

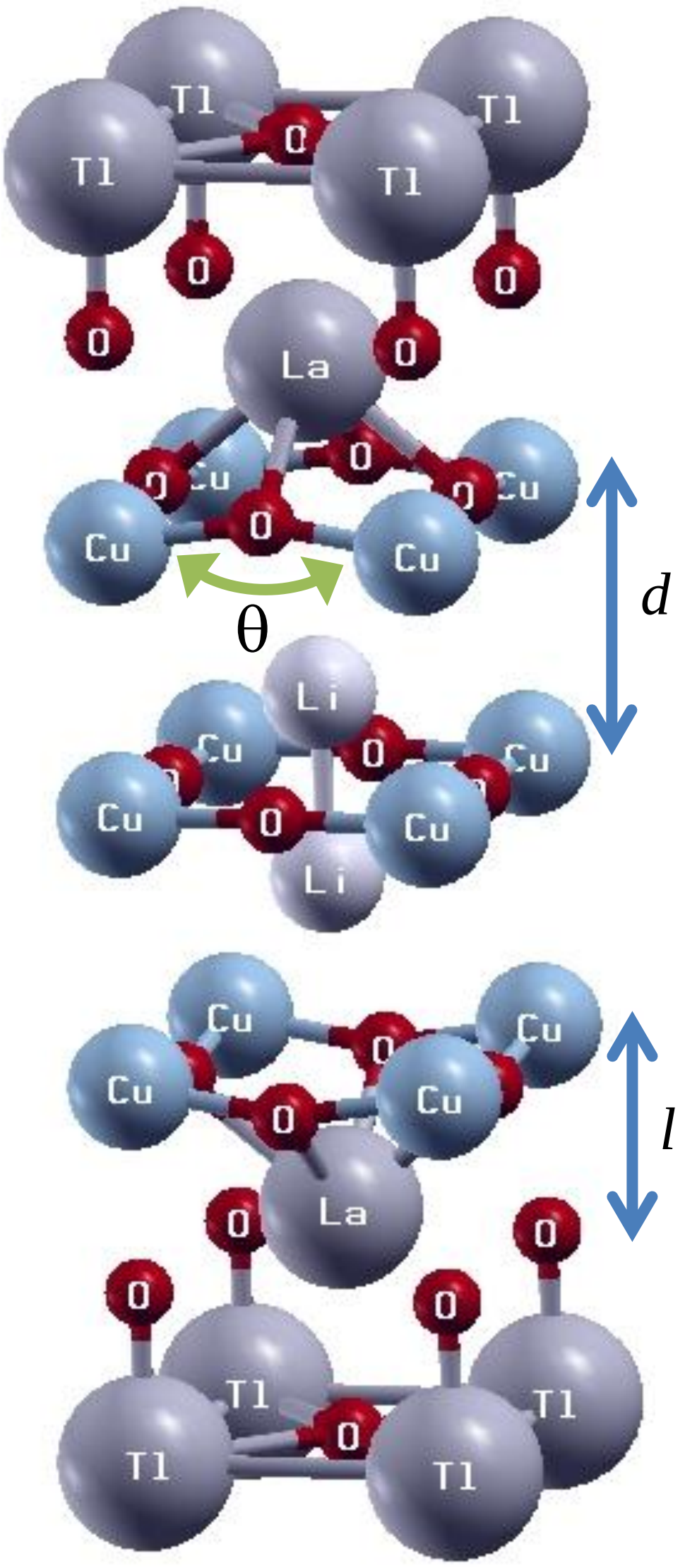
Tl
Tl
O
Tl
Tl
O
O
O
O
La
O
Cu
O
Cu
O
Cu
Cu
θ
d
Li
Cu
O
Cu
Cu
O
Cu
Li
Cu
O
Cu
Cu
O
Cu
l
La
O
O
O
O
O
Tl
O
Tl
Tl
Tl

**Figure 2**

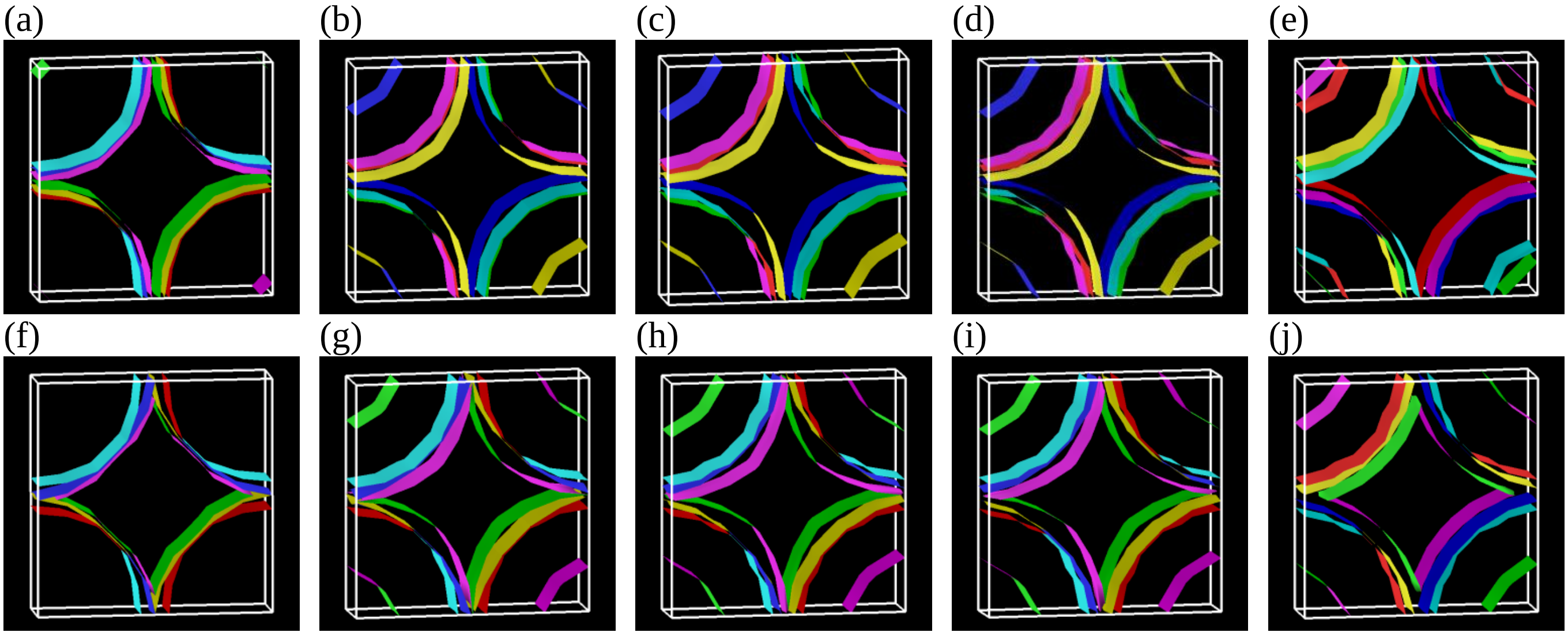

**Figure 3** (10-column figure)

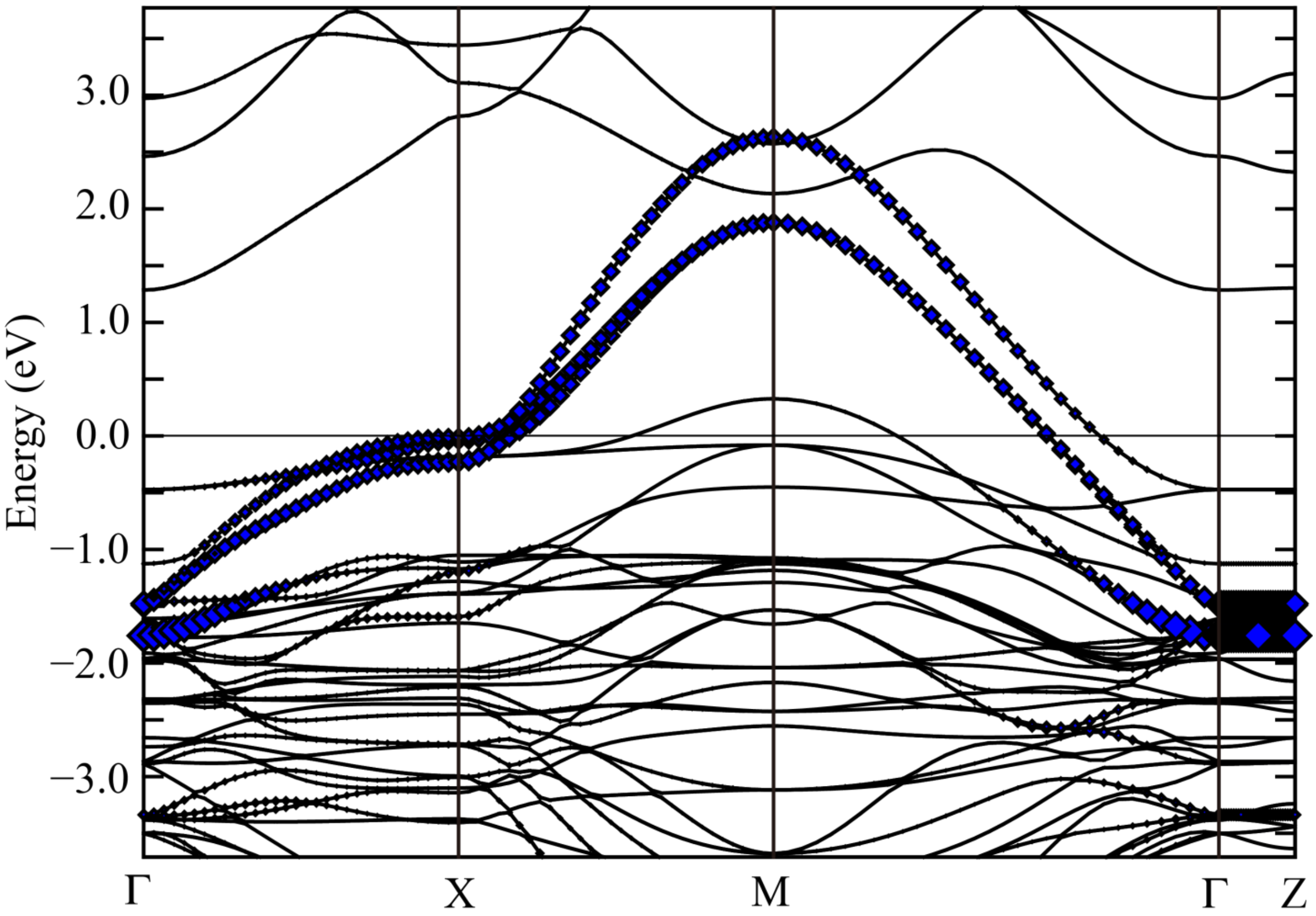

**Figure 4 (a)**

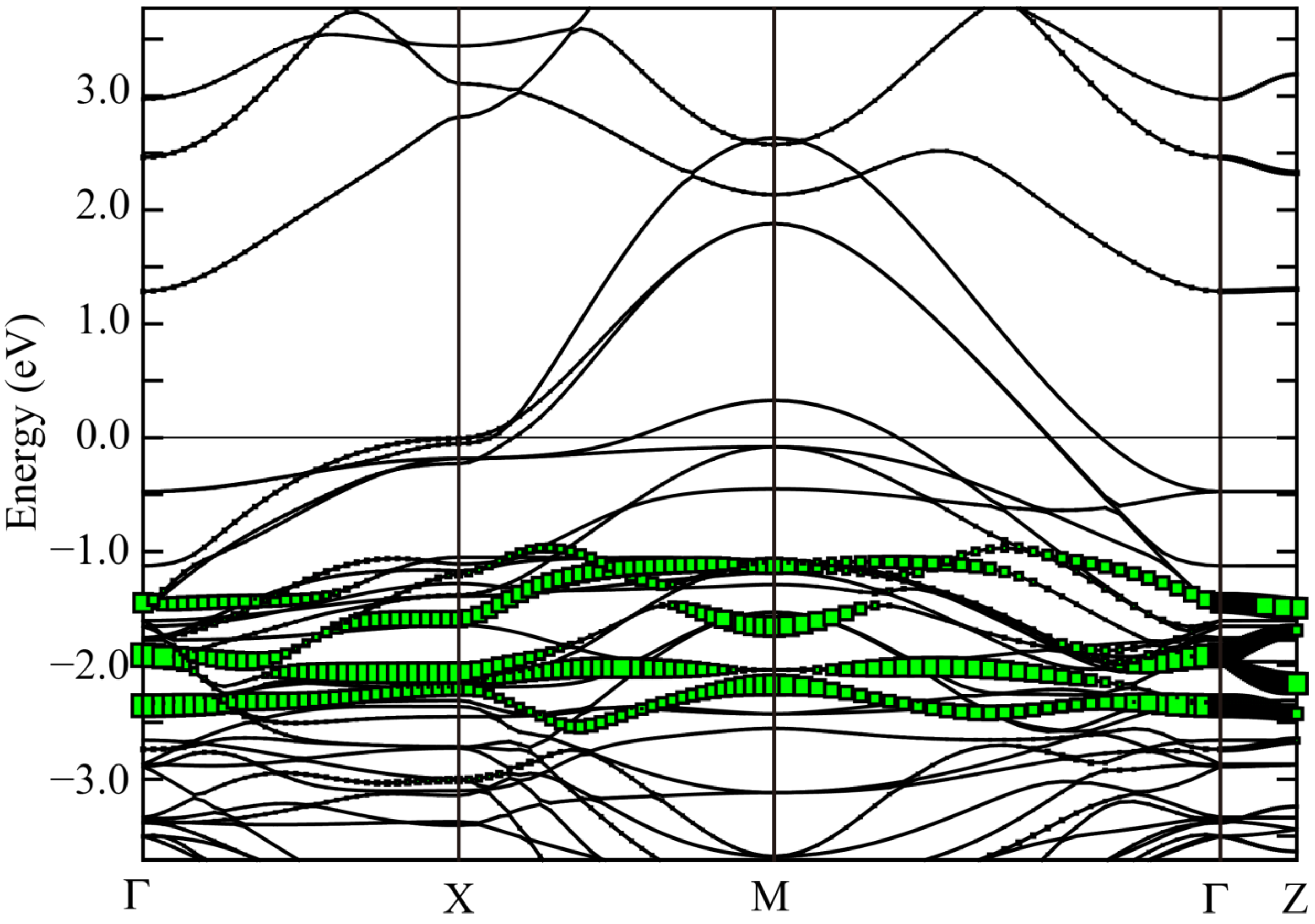

Figure 4 (b)

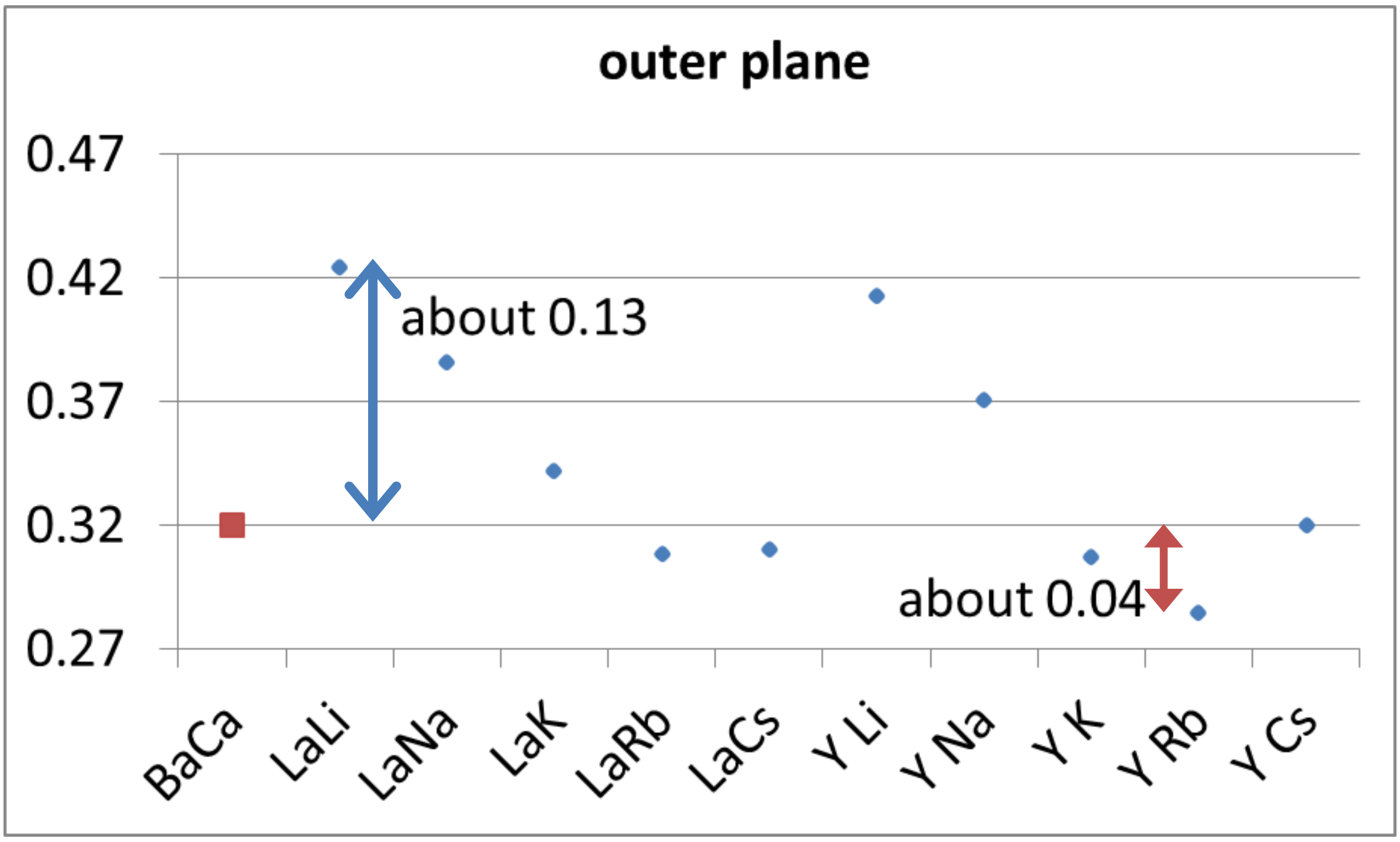

Figure 5

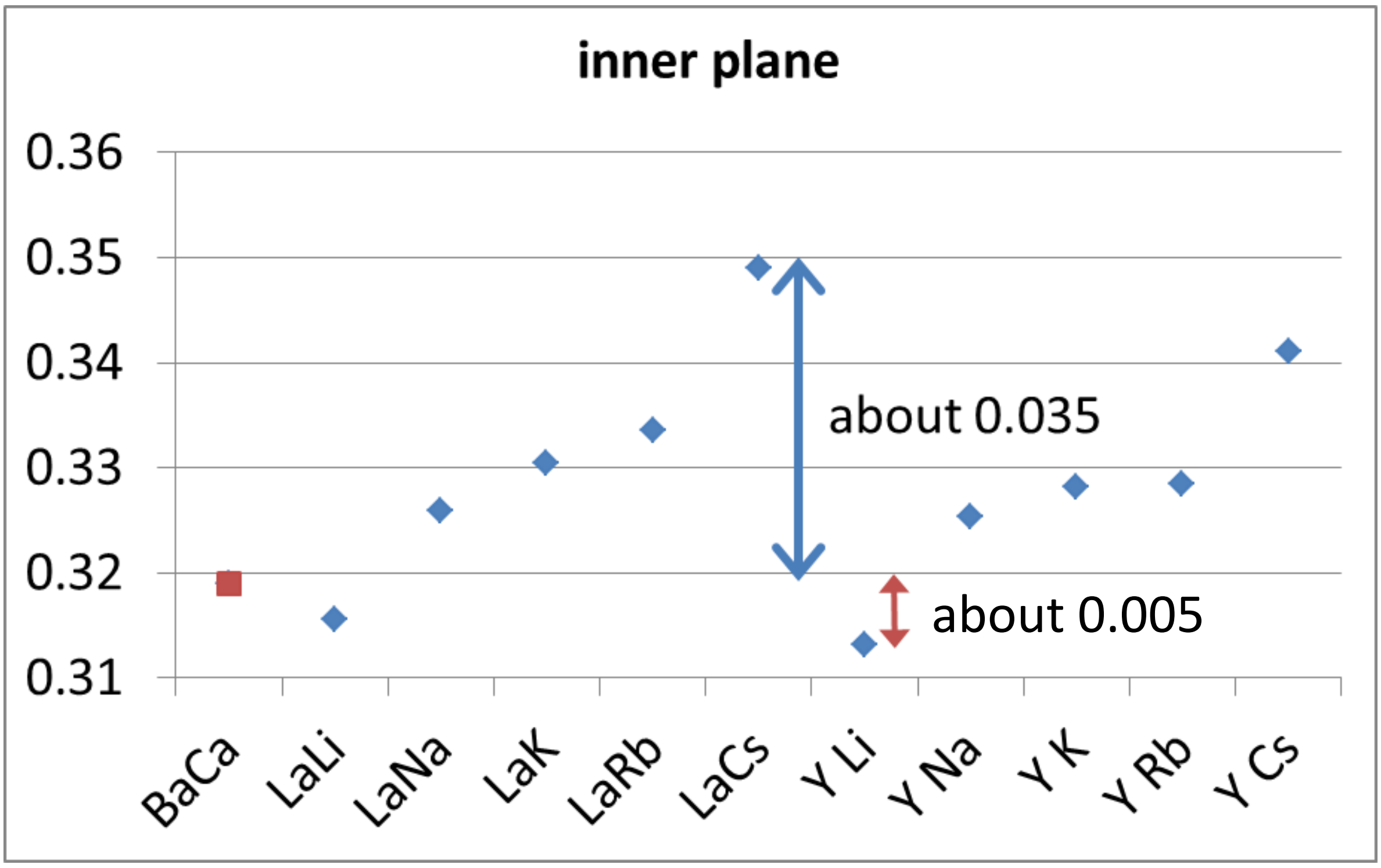

**Figure 6**

## Tables

| | Obs. | Calc. (cell fixed) | Calc. (cell opt.) |
|---|---|---|---|
| *a* [Å] | 3.8429 | 3.8429 | 3.853 |
| *c* [Å] | 15.871 | 15.871 | 16.18 |
| z at 2h site of Ba | 0.17197 | 0.16716 | 0.16659 |
| z at 2h site of Ca | 0.3961 | 0.3967 | 0.3968 |
| z at 2g site of Cu | 0.29842 | 0.29452 | 0.29583 |
| z at 2g site of O | 0.1258 | 0.1325 | 0.1303 |
| z at 4i site of O | 0.3024 | 0.3051 | 0.3058 |
| O-Cu-O angle, θ [° ] | 176.23 | 170.00 | 170.40 |
| Cu inter-plane distance, *d* [Å] | 3.20 | 3.26 | 3.30 |
| Cu-O distance *l* [Å] | 2.74 | 2.57 | 2.68 |

**Table 1**

| $TlX_2A_2Cu_3O_9$ | | | *A* | | | | |
|---|---|---|---|---|---|---|---|
| | | | Li | Na | K | Rb | Cs |
| O-Cu-O angle, | *X* | La | 161.89 | 164.83 | 167.20 | 167.22 | 166.41 |
| θ [° ] | | Y | 162.15 | 164.30 | 165.57 | 164.94 | 163.25 |
| Cu inter-plane distance, | *X* | La | 3.34 | 3.39 | 3.63 | 3.74 | 3.84 |
| *d* [Å] | | Y | 3.60 | 3.61 | 3.78 | 3.87 | 3.95 |
| Cu-O distance | *X* | La | 2.52 | 2.48 | 2.28 | 2.20 | 2.13 |
| *l* [Å] | | Y | 2.30 | 2.29 | 2.16 | 2.10 | 2.05 |

**Table 2**

| Material | Plane | $t_1$[eV] | $t_2$[eV] | $t_3$[eV] |
|---|---|---|---|---|
| $TlBa_2Ca_2Cu_3O_9$ | Outer | -0.5109 | 0.09383 | -0.06977 |
| | Inner | -0.5378 | 0.09531 | -0.07627 |
| $TlLa_2Li_2Cu_3O_9$ | Outer | -0.3917 | 0.08917 | -0.07712 |
| | Inner | -0.5424 | 0.09214 | -0.07905 |
| $TlLa_2Na_2Cu_3O_9$ | Outer | -0.4072 | 0.08492 | -0.07236 |
| | Inner | -0.5387 | 0.09223 | -0.08333 |
| $TlLa_2K_2Cu_3O_9$ | Outer | -0.4252 | 0.08004 | -0.06537 |
| | Inner | -0.5414 | 0.09254 | -0.08635 |
| $TlLa_2Rb_2Cu_3O_9$ | Outer | -0.4400 | 0.08257 | -0.05308 |
| | Inner | -0.5420 | 0.09306 | -0.08768 |
| $TlLa_2Cs_2Cu_3O_9$ | Outer | -0.4320 | 0.07941 | -0.05454 |
| | Inner | -0.5435 | 0.09579 | -0.09388 |
| $TlY_2Li_2Cu_3O_9$ | Outer | -0.3962 | 0.08214 | -0.08137 |
| | Inner | -0.5416 | 0.09201 | -0.07754 |
| $TlY_2Na_2Cu_3O_9$ | Outer | -0.4203 | 0.07536 | -0.08045 |
| | Inner | -0.5485 | 0.09527 | -0.08321 |
| $TlY_2K_2Cu_3O_9$ | Outer | -0.4253 | 0.08266 | -0.04804 |
| | Inner | -0.5406 | 0.09238 | -0.08504 |
| $TlY_2Rb_2Cu_3O_9$ | Outer | -0.4478 | 0.08188 | -0.04576 |
| | Inner | -0.5418 | 0.09238 | -0.08559 |
| $TlY_2Cs_2Cu_3O_9$ | Outer | -0.3997 | 0.07067 | -0.05733 |
| | Inner | -0.5410 | 0.09428 | -0.09018 |

**Table 3**